\documentclass[12pt]{article}
\usepackage{amsmath}
\usepackage{amssymb}
\usepackage{latexsym}
\usepackage{pb-diagram}

\newtheorem{thm}{Theorem}[section]
\newtheorem{prop}[thm]{Proposition} 
 \newtheorem{dfn}[thm]{Definition}
 \newtheorem{rmk}[thm]{Remark}
 
\newtheorem{conj}[thm]{Conjecture}
\newcommand {\pf}{\noindent{\bf Proof.}\ }
\newcommand{\complex}{{\mathbb C}}
\newcommand{\naturals}{{\mathbb N}}
\newcommand{\reals}{{\mathbb R}}
\newcommand{\torus}{{\mathbb T}}
\newcommand{\integers}{{\mathbb Z}}

\newcommand{\Hom}{{\rm Hom}}

\newcommand{\cala}{{\cal A}}

\newcommand{\calf}{{\cal F}}

\newcommand{\calo}{{\cal O}}

\newcommand{\calq}{{\cal Q}}
\newcommand{\cals}{{\cal S}}
\newcommand{\calt}{{\cal T}}

\newcommand{\calw}{{\cal W}}

\begin{document}

\title{A note on Q-algebras and quantization}
\author{Xiang Tang }


\maketitle

\begin{abstract}
In this note, we study Schwarz's conjecture on application of
Q-algebras to strict quantization. We prove that in the case of a
torus with a constant Poisson structure, Schwarz's formalism gives
the same star product as Rieffel \cite{rif:quantization}. We
construct twisted Fock modules as examples of quantization
dg-modules in the case of a compact K\"ahler manifold. In
particular, we relate this construction on $\complex\mathbb{P}^1$ to
 representations of a fuzzy sphere.
\end{abstract}
\section{Introduction}
The notion of a Q-algebra was introduced by Albert Schwarz
\cite{sc:ncsusy} in the study of noncommutative supergeometry and
gauge theory.

The idea of noncommutative geometry goes back to algebraic geometry
and the famous Gelfand-Naimark theorem. All geometric properties of
a closed smooth manifold $X$ are encoded in the algebra
$C^{\infty}(X)$ of smooth functions on $X$, and a map from $X$ to
another closed smooth manifold  $Y$ defines an algebra homomorphism
from $C^{\infty}(Y)$ to $C^{\infty}(X)$. This correspondence between
geometric spaces and commutative algebras was extended to
noncommutative geometry and algebras by Alain Connes and his
followers \cite{c:book}. Roughly speaking, in noncommutative
geometry, every associative algebra is considered as an algebra of
functions on a ``noncommutative space.'' Noncommutative geometry
studies the generalized notions of conventional geometry, and uses
our geometric intuition to obtain a new and better understanding on
some noncommutative (and even commutative) algebras, e.g. the
algebras of functions on noncommutative tori.

One can say (not very precisely) that supergeometry is
supercommutative noncommutative geometry. The algebra
$C^{\infty}(X)$ of smooth functions on a supermanifold $X$ is a
$\integers_2$-graded supercommutative algebra and a map between
supermanifolds $X$ to $Y$ is an algebra homomorphism from
$C^{\infty}(Y)$ to $C^{\infty}(X)$.  In general, we can ``define" a
``noncommutative supermanifold" as a ``space" having a
$\integers_2-$graded associative (but not necessary
supercommutative) algebra of functions. Of course, this definition
is as formal as the definition of noncommutative space in terms of
an associative algebra of functions. It is safe to say that
noncommutative supergeometry combines ideas of  both noncommutative
geometry and supergeometry.

A vector field on a supermanifold $X$ is a derivation on the
supercommutative algebra $C^{\infty}(X)$. An odd vector field
corresponds to an odd derivation, which changes the parity of an
element in $C^{\infty}(X)$.
\begin{dfn}
\label{dfn:q-mfld}A supermanifold equipped with an odd vector field
$Q$ obeying $Q^2=0$ is called a $Q-$manifold.
\end{dfn}

A natural generalization of Definition \ref{dfn:q-mfld} to
noncommutative geometry leads to the following definition.
\begin{dfn}
\label{dfn:q-alg} A Q-algebra $A$ is a $\integers_2$ (or
$\integers$) graded associative algebra with a derivation $Q$ of
degree 1 and an element $\omega\in A$ of degree 2 satisfying
\[
Q^2x=[\omega, x],
\]
for all $x\in A$.

We call $\omega$ the curvature of the Q-algebra $A$.
\end{dfn}

Schwarz \cite{sc:ncsusy, sc:deformation} studied the theory of
modules over a Q-algebra and proved a general duality theorem in
this framework which generalized the $SO(d,d;\integers)$ duality of
gauge theories on noncommutative tori to Q-algebras. In particular,
he discovered an interesting connection between $Q-$algebras and
deformation quantization.

The problem of formal deformation quantization of a general Poisson
manifold was solved by Kontsevich \cite{konts:deformation}.
Kontsevich's construction implicitly used the topological BV
sigma-model \cite{aksz:bv}, which was formulated by Cattaneo and
Felder \cite{cf:poi-sigma}. However, in Kontsevich's star product,
the parameter $\hbar$ is a formal one. The problem of whether one
can make $\hbar$ into a real number and construct a continuous
family of associative multiplications is much harder.  Using
Q-algebras, Schwarz \cite{sc:deformation} suggested a way that
allows one to circumvent the above strict quantization problem. We
explain this idea in the following paragraphs.

We recall Fedosov's method of formal deformation quantization of
symplectic manifolds. One starts with a standard formal Weyl algebra
$W$ corresponding to a symplectic vector space, and constructs a
Weyl algebra bundle $\calw$ over a symplectic manifold $X$ through
the principal symplectic group bundle. A torsion free symplectic
connection on $X$ lifts to a connection $\nabla$ on bundle $\calw$,
which determines a structure of a Q-algebra on the algebra of
$\calw-$valued differential forms $\Omega(X, \calw)$ with
curvature\footnote{The curvature of a right module $M$ over a
Q-algebra $(A, Q, \omega)$ with a connection $\nabla$ is defined to
be $\nabla^2+\hat{\omega}\in \Hom(M)$, where $\hat{\omega}$ is the
right multiplication of $\omega$. } equal to $R$. Fedosov's results
on deformation quantization can be reformulated into the property
that the Q-algebra $(\Omega(X, \calw), \nabla, R)$ is equivalent to
a new Q-algebra $(\cala, D, \nu)$ such that the curvature $\nu$ is
equal to $-\frac{i}{\hbar}\omega+\tilde{\omega}$, where $\omega$ is
the symplectic form on $X$, and $\tilde{\omega}$ is a 2-cocycle in
$\Omega^2(X, \complex[[\hbar]])$. (Two Q-algebras are equivalent if
they are isomorphic and their derivations are related by the formula
$Q'=Q+[\gamma,\ ]$, for $\gamma \in A$). This equivalent Q-algebra
can be viewed as a differential graded algebra $(\cala', D)$ because
its curvature $\nu$ is a center element. It is proved by Fedosov
\cite{fe:book} that the differential graded algebra $(\cala', D)$ is
quasi-isomorphic to an associative algebra $A(X)$ that can be
considered as a quantization of $X$. (Notice that $A(X)$ is
isomorphic to $C^{\infty}(X)[[\hbar]]$ as a vector space, and the
product on $A(X)$ is translated to a star product on
$C^{\infty}(X)[[\hbar]]$.) Since $(\cala', D)$ is quasi-isomorphic
to $A(X)$, they share the same geometric properties, i.e.
cohomology, characteristic classes, and K-theory, etc. In
particular, the category of projective modules of $A(X)$ is
equivalent to the category of projective dg-modules of $(\cala', D)$
up to quasi-isomorphisms. (A dg-module over a differential graded
algebra is a module with a flat connection.) It is easy to check
that there is a one-to-one correspondence between projective
dg-modules of the differential graded algebra $(\cala', D)$ and
projective modules of the Q-algebra $(\cala, D, \nu)$ with curvature
equal to the multiplication of $\nu$. By the equivalence between the
Q-algebras $(\Omega(X, \calw), \nabla, R)$ and $(\cala, D, \nu )$,
there is again a one-to-one correspondence between the projective
modules with the fixed curvature $\hat{\nu}$. We conclude that
projective modules of $A(X)$, the deformation quantization of $X$,
are related to projective modules of $(\Omega(X,\calw), \nabla, R)$
with curvature equal to $\hat{\nu}$. We emphasize that it is
important to work with the Q-algebra $(\cala, D, \nu)$ instead of
just the differential algebra $(\cala', D)$.

Schwarz's  crucial observation \cite{sc:deformation} is that the
parameter $\hbar$ in $\Omega(X, \calw)$ can be made into a real
number, so that $\Omega(X, \calw)$ becomes a complete topological
algebra. This is achieved by an appropriate modification of the
notion of Weyl algebra, which becomes a complete nuclear local
convex algebra depending smoothly on the real parameter $\hbar$.
Then one can consider a ``new" Weyl algebra bundle $\calw$, and a
``topological" $Q$-algebra $(\Omega(X, \calw), \nabla, R)$ as the
space of sections of $\calw$. It is possible that one can find an
equivalent Q-algebra $(\cala, D, \nu)$ such that the curvature
$\Omega$ is in the center of $\cala$, and the associated
differential graded algebra $(\cala', D)$ is quasi-isomorphic to an
associative topological algebra $A(X)$, which is isomorphic to
$C^\infty(X)$ as a vector space. Then the algebra $A(X)$ can be
considered as a strict deformation quantization of $X$. We should
remark that it is very hard (maybe impossible) to construct a
topological Q-algebra $(\cala, D, \nu)$ satisfying the above
requirements. However, Schwarz \cite{sc:deformation} suggested that
instead of searching for such an equivalent differential graded
algebra, one should study the Q-algebra $(\Omega(X, \calw), \nabla,
R)$ directly. He conjectured that the Q-algebra $(\Omega(X, \calw),
\nabla, R)$ can be considered as a replacement of a convergent
deformation quantization. In particular, if a strict deformation
quantization of $X$ exists, the category of projective modules of
the Q-algebra $(\Omega(X, \calw), \nabla, R)$ with curvature equal
to $\nu$ should be related to the category of projective modules of
the strict deformation quantization of $X$. Schwarz conjectured that
if the strict deformation quantization does not exist, the above
category of modules should be regarded as a replacement of the
category of projective modules of the strict deformation
quantization. In any case, by studying this Q-algebra and its
modules, we may obtain the information of the strict deformation
quantization of $X$. Furthermore, this study may also lead to some
``no go" theorems in strict deformation quantization, which should
be very interesting.

The results in this paper have two important improvements with
respect to Schwarz's original work \cite{sc:deformation}. Firstly,
to realize the above proposal, we need to know how to modify the
formal Weyl algebra into a ``nice" topological algebra. This
question has many different possible answers. Schwarz
\cite{sc:deformation} suggested one candidate for the answer. We
find that the proposed algebra is hard to work with in the case of
noncommutative tori. Gracia-Bondia and V\'arilly
\cite{gv:distribution} studied the largest possible $*$-subalgebra
of tempered distributions on $\reals^{2n}$ where the ``twisted
product"(convolution product) is defined. This $*$-subalgebra is
invariant under Fourier transform and is very ``big", and contains
rapidly decreasing smooth functions, distributions of compact
support and all polynomials, etc. Later, Dubois-Violette, Kriegl,
Maeda and Michor \cite{dkmm:smooth-algebra} considered a smaller
subalgebra $\calo_M'(\reals^{2n})$ of speedily decreasing
distributions on $\reals^{2n}$, whose definition will be recalled in
Section 2.1. They suggested that $(\calo_M'(\reals^n),
\hat{\ast}_{\hbar})$ be viewed as the covering space of a
noncommutative torus $\calt_{\theta}$, where $\theta$ is a constant
Poisson structure on an $n-$dimensional torus. In this paper, we
will follow this suggestion and use the subalgebra
$(\calo_C,\ast_{\hbar})$, the Fourier transform of
$(\calo_M'(\reals^n), \hat{\ast}_{\hbar})$, as the fiber of the new
Weyl algebra. The good property of the algebra $\calo_C$ is that if
$f\in \calo_C$ and $df=0$, then there is $g\in \calo_C$ with $dg=f$.
Therefore the de Rham differential is exact on $\calo_C$.

Secondly, we recall that in \cite{sc:deformation} a dg-module over a
Q-algebra $(A, Q, \omega)$ is defined to be an $A$-module $M$ with a
connection $\nabla_M$ such that the curvature
$\nabla_M^2+\hat{\omega}$ equals 0 as an endomorphism on $M$, where
$\hat{\omega}$ is the multiplication of $\omega$ on $M$. This is a
natural generalization of the notion of a dg-module over a
differential graded algebra. However, this definition is not very
useful for our purpose. As described above, in Fedosov's formal
deformation quantization of a symplectic manifold, projective
modules of the associative algebra $A(X)$ are related to projective
modules of the Q-algebra $(\Omega, \nabla, R)$ with curvature equal
to $\nu$. This experience suggests that we should allow a dg-module
over a Q-algebra to have a central curvature, not just 0. Therefore,
we introduce the following modified definition of a dg-module over a
$Q$-algebra.

\begin{dfn}
\label{dfn:dg-module} A dg-module over a Q-algebra $(A, Q, \omega)$
is a right $A$-module $M$ with a connection $\nabla_M$ such that the
curvature $\nabla^2_M+\hat{\omega}$ as an element in $End(M)$ is the
multiplication by a center element in $A$.

In the case of a symplectic manifold $(X, \omega)$, we call a
right $(\Omega(X, \calw), \nabla, R)$-module a quantization
dg-module if its curvature is equal to the multiplication by
$-\frac{i}{\hbar}\omega$.
\end{dfn}

The goal of this paper is to study Schwarz's conjecture
\cite{sc:deformation} on some examples. We prove that his conjecture
is correct in these cases. The paper is arranged as follows. In
Section 2, we study the case of a torus with a constant Poisson
structure, and construct a Q-algebra $(\Omega(\torus^n, \calw), Q)$
with curvature equal to $-\frac{i}{\hbar}\omega$. We show that,
viewed as a differential algebra, $(\Omega(\torus^n, \calw), Q)$ is
quasi-isomorphic to the algebra of the corresponding quantum torus.
By this quasi-isomorphism, we know that the algebra
$(\Omega(\torus^n, \calw), Q, -\frac{i}{\hbar}\omega)$ contains as
much information as a quantum torus. In Section 3, we will construct
a Fock module over $\Omega(X, \calw)$ on a compact K\"ahler
manifold, which leads to examples of quantization dg-modules over
$(\Omega(X, \calw), \nabla, R)$. We show that in the case of a
standard 2-sphere, our construction is connected to the well-known
fuzzy sphere. We end this note with a conjecture on quantization
dg-modules.

Acknowledgement: First of all, I would like to thank Albert Schwarz,
who should be considered as a ``virtual co-author'', for many
interesting questions and helpful suggestions. I also want to thank
Alan Weinstein for helpful conversations.

\section{Quantum Tori}
\label{torus} In this section, we exhibit a construction for
quantizing a constant Poisson torus following the idea described
in the Introduction. We obtain the quantization in three steps:
\begin{enumerate}
\item quantization of $\reals^n$; \item $Q-$algebra; \item quantization.
\end{enumerate}
\subsection{Quantization of $\reals^n$}
In this part, following the ideas of Dubois-Violette, Kriegl, Maeda
and Michor \cite{dkmm:smooth-algebra}, we introduce an algebra that
plays the role of the Weyl algebra in the formal deformation
quantization.

We recall here some well-known results in the distribution theory.

Let $\cals(\reals^n)$ be the space of rapidly decreasing smooth
functions $f$ for which $x\mapsto (1+|x|^2)^k\partial ^{\alpha}f(x)$
is bounded for all $k\in \naturals$ with all multi-indices
$\alpha\in \naturals _0^{n}$.  We endow it with the locally convex
topology described by these conditions, so that it is a nuclear
Fr\'echet space. Let $\cals '(\reals ^n)$ be the dual space of
$\cals(\reals ^n)$ consisting of tempered distributions.

Let $\calo_C(\reals ^n)$ be the space of all smooth functions $f$ on
$\reals ^n$ for which there exists $k\in \integers$ such that
$x\mapsto (1+|x|^2)^k \partial ^{\alpha}f(x)$ is bounded for each
multi-index $\alpha \in \naturals _0^n$. We endow this space with
the locally convex topology. $\calo_C(\reals^n)$ is a complete
nuclear (LP-) space. Let $\calo'_C(\reals ^n)$ be its dual space
consisting of rapidly decreasing distributions.

Let $\calo_M (\reals ^n)$ consists of all smooth functions on
$\reals^n$ such that for each multi-index $\alpha \in \naturals _0
^n$, there exists $k\in \integers$ such that $x\mapsto
(1+|x|^2)^k\partial ^k f(x)$ is bounded. We endow it with the
locally convex topology described by this condition. $\calo_M$ is
usually called the space of tempered smooth functions, whose dual
space $\calo'_M (\reals ^n)$ is called the space of speedily
decreasing distributions.

We have the following relations among the above introduced spaces:
\[
\begin{diagram}
\node{\cals}\arrow{e,t,T}{\iota}\arrow{s,l,T}{\calf}\node{\calo_C}
\arrow{e,t,T}{\iota}\arrow{s,l,T}{\calf}\node{\calo_M}\arrow{e,t,T}
{\iota}\arrow{s,l,T}{\calf}\node{\cals'}\arrow{s,l,T}{\calf}\\
\node{\cals}\arrow{e,t,T}{\iota
^*}\node{\calo_M'}\arrow{e,t,T}{\iota^*}\node{\calo_C'}\arrow{e,t,T}{\iota^*}\node{\cals}
\end{diagram}
\]
where in the above, $\iota$ is the embedding map, and $\calf$ is
the Fourier transform.

In Section 3, \cite{dkmm:smooth-algebra}, the following theorem is proved.
\begin{thm}
\label{thm:dfn-moyal-prod}The following convolution
\begin{equation}
\label{eq:star-dfn} (a\hat{\ast}_{\hbar}b)(x)=
\int_{\reals^{2}}a(x-y)b(y)e^{-\frac{i\hbar}{2}\omega(x,y)}{\rm
d}y
\end{equation}
where $\omega$ is the standard symplectic two form on $\reals^{2}$,
defines a star product $\hat{\ast}_{\hbar}$. This product is an
associative bounded multiplication on the space
$\calo_M'(\reals^{2})$ of speedily decreasing distributions. It is
smooth in the variable $\hbar \in \reals$.

The following formula
\begin{equation}\label{eq:rep-star}
\hat{a}\stackrel{def}{=}\int_{\reals^2}a(t,
s)e^{\frac{i\hbar}{2}ts}e^{itQ}e^{isP}{\rm d}s{\rm d}t,
\end{equation} defines a bounded linear mapping $\calo'_M(\reals
^2)\to L(\cals(\reals), \cals(\reals))$, where
$e^{isP}f(u)\stackrel{def}{=}f(u+s\hbar)$ and
$e^{itQ}f(u)\stackrel{def}{=}e^{itu}f(u)$. This representation is
injective if $\hbar\ne0$, and is an algebra homomorphism from the
product (\ref{eq:star-dfn}) to the operator composition. The analog
on $\reals ^{2n}$ also holds.
\end{thm}

We calculate the pullback of $\hat{\ast}_{\hbar}$ to $\calo_C$
through Fourier transformation. Let $f,g\in \calo_C$. Then
\[
\begin{split}
&\calf^{-1}(\calf(f)\hat{\ast}_{\hbar}\calf(g))(x)\\
=&\frac{1}{(2\pi)^{2n}}
\int_{\reals^{2n}}e^{i x\cdot y}\calf(f)\hat{\ast}_{\hbar}\calf(g)(y){\rm d}y\\
=&\frac{1}{(2\pi)^{2n}}\int_{\reals^{2n}}e^{i x\cdot y}\int_{\reals^{2n}}\calf(f)(y-u)\calf(g)(u)e^{-\frac{i\hbar}{2}\omega(y,u)}{\rm d}u{\rm d}y\\
=&\frac{1}{(2\pi)^{2n}}\int_{\reals^{2n}}\int_{\reals^{2n}}e^{ix\cdot y-\frac{i\hbar}{2}\omega(y,u)}{\rm d}y{\rm d}u\int_{\reals^{2n}}e^{-i(y-u)\cdot w}f(w){\rm d}w
\int_{\reals^{2n}}e^{-iu\cdot z}g(z){\rm d}z\\
=&\frac{1}{(2\pi)^{2n}}\int_{\reals^{2n}}\int_{\reals^{2n}}\int_{\reals^{2n}}\int_{\reals^{2n}}e^{ix\cdot y-\frac{i\hbar}{2}\omega(y,u)-i(y-u)\cdot w
-i u\cdot z}f(w)g(z){\rm d}y{\rm d}u{\rm d}w{\rm d}z\\
=&\int_{\reals^{2n}}\int_{\reals^{2n}}f(w)g(z)e^{\frac{-2i}{\hbar}\omega(x-z,
x-w)}{\rm d}z{\rm d}w.\\
\end{split}
\]

Therefore, we define on $\calo_C(\reals^{2n})$ a product $\ast_\hbar$ by
\begin{equation}\label{eq:star} f\ast_\hbar
g(x)=\frac{1}{(\pi
\hbar)^{2n}}\int_{\reals^{2n}}\int_{\reals^{2n}}f(x+u)g(x+v)e^{\frac{-2i}{\hbar}\omega(u,v)}{\rm
d}u{\rm d}v.
\end{equation}

In the following, we will use $(\calo_C(\reals ^{2n}),
\ast_{\hbar})$, a complete nuclear (LP)-algebra, as our model for a
noncommutative euclidean space.

\subsection{Construction of a Q-algebra} In this subsection,
we will construct a Q-algebra, actually a differential graded
algebra. (We say that a constant symplectic (Poisson) torus is a
torus with a constant symplectic (Poisson) structure.) For
convenience, we work explicitly on a 2-dimensional torus; the same
method can be generalized to an $n$-dimensional torus without extra
effort.

Let $(\torus ^2, \omega)$ be a 2-dimensional symplectic torus and
$C^{\infty} (\torus ^2)$ be the algebra of smooth functions on
$\torus ^2$. Denote $\calw$ to be a trivial infinite-dimensional
vector bundle over $\torus^2$ with fiber isomorphic to
$\calo_C(\reals ^2)$. We consider the following algebra
\[
\Omega(\torus^2, \calw)\stackrel{def}{=}\Gamma^{\infty}(\wedge ^*
\torus ^2)\boxtimes \calo_C(\reals ^2),
\]
where $\boxtimes$ is the projective tensor product with respect to
the topologies. We remark that we need to replace the standard
symplectic form in (\ref{eq:star}) by the symplectic form on the
torus, in order to include the information of the symplectic
(Poisson) torus in the definition of $\ast_{\hbar}$ on $\calo_C$.

The product on $\cala$ is defined as
\[
f\ast_{\hbar} g(x,y)\stackrel{def}{=}(f(x)\ast_{\hbar} g(x))(y),
\]
where $x$ and $y$ are coordinates on $\torus^2$ and $\reals^2$,
$f(x)$ and $g(x)$ are restrictions of $f$ and $g$ to $x$, and the
$\ast$ between $f(x)$ and $g(x)$ is the `twisted product' defined in
Theorem \ref{thm:dfn-moyal-prod}. $\Omega(\torus^2, \calw)$ can be
viewed as the space of smooth sections of $\calw$, and is naturally
graded by the degree of differential forms. In the following, we
will define a $Q$ operator on $\Omega(\torus^2, \calw)$.

We define
\[
Q\stackrel{def}{=}dx^i\wedge(\frac{\partial}{\partial x^i}- \frac
{\partial}{\partial y^i}),
\]
where $x^i$s are the coordinates on $\torus^2$ and $y^i$s are the
coordinates on $\reals^2$.

It is straightforward to check that $deg(Q)=1$ and $Q^2=0$.
Therefore, $(\Omega(\torus^2, \calw), Q)$ is a Q-algebra with
$\omega=-\frac{i}{\hbar}dx^1\wedge dx^2 $. Since $\omega$ is a
center element in $\cala$ and therefore we can view
$(\Omega(\torus^2, \calw), Q)$ as a differential graded algebra.

\subsection{Quantization}
In this part, we prove that the Q-algebra $\Omega(\torus^2, \calw)$
defined in the previous subsection is quasi-isomorphic to
$C^{\infty}(\torus^2)$ as differential graded algebras. By
identifying the 0-th cohomologies of $(\cala, Q)$ and $C^{\infty}
(\torus ^2)$, we will obtain a quantization map.

\begin{thm}
\label{thm:acyclic} The differential algebra $(\Omega(\torus^2,
\calw), Q)$ is acyclic and has 0-th cohomology equal to
$C^{\infty}(\torus ^2)$.
\end{thm}
$\pf$The crucial part of the proof already appears in
\cite{dkmm:smooth-algebra} for different purposes. By a change of
coordinates $\tilde{x}=x-y,\ \tilde{y}=x+y$, we have the following
observations.
\begin{enumerate}
\item $\Omega(\torus^2, \calw)=\Gamma^{\infty}(\wedge^*
\torus ^2)\boxtimes \calo_C(\reals^2)$ is isomorphic to
$\Gamma^{\infty}(\wedge^*_-(\torus^2\times \reals^2))\boxtimes
\calo_C(\reals^2)$, where $\wedge_- ^*(\torus^2\times \reals^2)$
denotes the differential forms on $\torus^2\times \reals^2$
along the direction of $\tilde{y}$.
\item Under the above isomorphism, $Q$ is mapped to $\tilde{Q}=d\tilde{y} \wedge \frac{\partial}{\partial
\tilde{y}}$.
\end{enumerate}

The following are reasons for the acyclicity of $\tilde{Q}$.
\begin{enumerate}
\item Poincar\'e lemma for $\tilde{Q}$. It is easy to see that
$\tilde{Q}$ is the de Rham differential along the $\tilde{y}$
direction on $\torus^2\times \reals^2$, and subspace of
$\torus^2\times \reals^2$ along the $\tilde {y}$ direction is
isomorphic to $\reals^2$. Therefore, the Poincar\'e lemma is true
on $\tilde{Q}$.
\item The homotopy formula constructed in the Poincar\'e
lemma, i.e. $\psi(x)=\int_0 ^1\sum_i \phi_i(tx)x_idt$ for
$\phi=\sum_i \phi_i$ and $d\phi=0$. We know that $\phi_i\in
\calo_C(\reals^2)$ implies $\psi\in \calo_C(\reals^2)$ by a
simple estimation. (This has been observed in
\cite{dkmm:smooth-algebra}.)
\end{enumerate}

Therefore, by the isomorphism between $Q$ and $\tilde{Q}$, we conclude that $Q$ is also acyclic.
The 0-th cohomology of $Q$ is equal to the solution of the
following equation in $\Omega(\torus^2, \calw)$:
\begin{equation}
\label{eq:solution} (\frac{\partial}{\partial
x^i}-\frac{\partial}{\partial y^i})f=0.\ \ \ i=1,\ 2.
\end{equation}

Obviously, the solutions of (\ref{eq:solution}) are functions on
$\torus^2\times \reals^2$ of the form $g(x+y)$. Therefore, the map
$\sigma: H^0(\Omega(\torus^2, \calw), Q)\to C^{\infty}(\torus)$,
$\sigma(g)=g|_{y=0}$, defines an isomorphism between the 0-th
cohomologies of $Q$ and $C^{\infty}(\torus^2)$.$\Box$

By Theorem \ref{thm:acyclic}, we define a quantization map
$\calq=\sigma^{-1}:C^{\infty}(\torus^2)\to H^0(\Omega(\torus^2,
\calw), Q)$ by $\calq(f)(x, y)\stackrel{def}{=}f(x+y)$. A star
product on $C^{\infty}(\torus^2)$ can be defined as
\[
f\star_\hbar g(x)\stackrel{def}{=}\sigma(\calq(f)\ast_{\hbar}
\calq(g)).
\]
By construction, $\star_\hbar$ is smooth with respect to $\hbar$.

According to Theorem \ref{thm:acyclic} and the above construction
of $\star_\hbar$, we have the following proposition.
\begin{prop}
\label{prop:quant} If we view the algebra $(C^\infty(\torus^2),
\star_\hbar)$ as a differential graded algebra with $0$
differential and only nontrivial 0-degree component, the map
$\calq:(C^\infty(\torus^2), \star_\hbar, 0) \to (\Omega(\torus^2,
\calw), \ast_\hbar, Q)$ is a quasi-isomorphism.
\end{prop}

We can look at the bundle $V$ on $\torus^2$,
$V=\cals(\reals)\times \torus^2\to \torus^2$. By Theorem
\ref{thm:dfn-moyal-prod}, there is a natural representation of
$\calo_C$ on $\cals(\reals)$. Hence, we obtain an action of
$\Omega(\torus^2, \calw)$ on $V$ acting by pointwise
multiplication. Accordingly, we have a representation of
$(C^{\infty}(\torus^2), \star_{\hbar})$ by composing $R$ with
$\calq$. By this representation, we can define an involution
$^{\ast}$ and a pre $C^*-$norm on $(C^{\infty}(\torus^2),
\star_{\hbar})$ using Hilbert module techniques. It is easy to
check that $(C^{\infty}(\torus), \star_{\hbar}, ^{\ast_{\hbar}},
||\ ||_{\hbar})$ satisfies the definition of a strict deformation
quantization as defined in Rieffel \cite{rif:quantization}.

Therefore, by using $Q-$algebras, we have obtained a strict
deformation quantization of a Poisson torus.

\begin{rmk}
Our results of strict deformation quantization of a Poisson torus
actually coincides with Rieffel's \cite{rif:quantization}
construction. What we have done here can be considered as a
geometric explanation of Rieffel's method. Similarly, we can use
this Q-algebra method to obtain all the examples constructed in
Rieffel \cite{rif:quantization}.
\end{rmk}
\section{Fock module and dg-modules}
In this section, we construct some quantization dg-modules of the
Q-algebra $(\Omega(X, \calw), \nabla, R)$. We first introduce the
notion of a Fock module over the Q-algebra on a compact K\"ahler
manifold.

Let $X$ be a compact K\"ahler manifold with $\dim_\complex(X)=n$. At
any $x$ of $X$, the tangent space $T_xX$ is equipped with a
riemannian metric. We define an idempotent element $p_{\hbar}=2^n
exp(-\frac{|y|^2}{\hbar})$ in $(\calo_C(T_x X), \ast_\hbar)$, where
$|\cdot |$ denotes the riemannian metric. It is easy to check that
$p_{\hbar}\ast_{\hbar}p_{\hbar}=p_{\hbar}$, and according to
\cite{gv:distribution}, Equation (27), $p_\hbar \ast_\hbar
\bar{y}^i=0$ for $1\leq i\leq n$. Theorem 4, \cite{gv:distribution}
proves that $p_\hbar\ast_\hbar \calo_C$ is a subspace of
$\cals(\reals^{2n})$.

Furthermore $p_{\hbar}$ is a well defined function over $TX$ and
therefore defines a smooth section of $\calw$. We define a right
projective $\Omega(X, \calw)$ module $\calf$ by $p_\hbar
\ast_\hbar \Omega(X, \calw)$.

On a K\"ahler manifold, there is a natural choice of a symplectic
connection $\nabla$, the Levi-Civita connection, i.e. $\nabla=d+\Gamma$.
The operator $Q$ on $\Omega(X, \calw)$ is written as
\[
Q(a)=da+\frac{1}{\hbar}[\Gamma , a],\ \textit{for all}\ a\in
\Omega(X, \calw),
\]
where $\Gamma=\Gamma_{ij}\bar{y}^i y^j$ and
$\Gamma_{ij}=\bar{\Gamma}_{ji}$.

One can define a connection $\nabla_{\calf}$ on $\calf$ using $Q$ as
follows,
\[
\nabla_{\calf}(p_\hbar \ast_\hbar a)=p_\hbar\ast_\hbar Q(p_\hbar)\ast_\hbar
a+p_\hbar \ast_\hbar Q(a).
\]
It is straight forward to check that $Q(p_\hbar)=0$, so
\[
\nabla_\calf (p_\hbar \ast_\hbar a)=p_\hbar \ast_\hbar
Q(a)=p_\hbar\ast_\hbar d(a)+p_\hbar \ast_\hbar [\Gamma, a].
\]
Noticing that $p_\hbar \ast_\hbar \bar{y}^i=0$, we have that
\begin{equation}
\label{eq:connection} \nabla_\calf(p_\hbar \ast_\hbar a)=p_\hbar
\ast (da -a \ast_\hbar
(\frac{1}{\hbar}\Gamma_{ij}\bar{y}^i\ast_\hbar y^j)).
\end{equation}

Let $\Omega$ be the curvature form of $\nabla$. And we compute the
curvature of $\nabla_\calf$ to be
\[
\nabla_\calf ^2(p_\hbar \ast_\hbar a)=p_\hbar \ast_\hbar
(-\frac{1}{\hbar}R),
\]
where $R$ is equal to $R_{ij}\bar{y}^i\ast y^j$.

We recall that $Q^2(a)=\frac{1}{\hbar}[R, a]$, and conclude that
\begin{prop}
\label{prop:fock}
The Fock module $(\calf, \nabla_\calf)$ has zero curvature.
\end{prop}

In the following, we want to twist this Fock module by a line bundle
to get a quantization dg-module over $(\Omega(X, \calw), \nabla,
R)$. We consider a rank one right projective module $H_m$ of
$C^\infty(X)$ with the curvature equal to $-\textrm{im }\omega$. We
identify $H_m$ with the space of $p_mC^\infty(X)$, where $p_m$ is a
projective matrix valued in $C^\infty(X)$. Define
$P_m\stackrel{def}{=}p_\hbar p_m$ in $\Omega(X, \calw)$. We notice
that $p_\hbar\ast_\hbar p_m=p_m\ast_\hbar p_\hbar=p_\hbar p_m$ and
$P_m\ast_\hbar P_m=P_m$. Therefore
$\calf_m\stackrel{def}{=}P_m\ast_\hbar \Omega(X, \calw)$ defines a
right projective module over $\Omega(X, \calw)$ with the connection
$\nabla_m$ equal to
\[
\nabla_m(P_m\ast_\hbar a)=P_m\ast_\hbar Q(a)+P_m\ast_\hbar
\nabla_{H_m}(p_m)\ast_\hbar a.
\]
By a calculation similar to that used in the case of the Fock
module, the curvature of $\calf_m$ is equal to $-\textrm{im
}\omega$. By Definition \ref{dfn:dg-module}, $\calf_m$ is a
quantization dg-module over $(\Omega(X, \calw), \nabla, R)$ when
$\hbar=\frac{1}{m}$.

\begin{prop}
\label{prop:tw-fock} The twisted Fock module $\calf_m$ is a
quantization dg-module over $(\Omega(X, \calw),$ $ \nabla, R)$
when $\hbar=\frac{1}{m}$.
\end{prop}

We recall the procedure for the Berezin-Toeplitz quantization of
$\complex \mathbb{P}^1$. One considers the hyperplane bundle $H$,
and its powers $H^{\otimes l}$ over $\complex \mathbb{P}^1$.
According to the Riemann-Roch theorem, for $l\geq 0$, the space
$B_{l}$ of holomorphic sections of $H^{\otimes l}$ is finite. One
considers the action of $C^\infty(\complex \mathbb{P}^1)$ on $B_l$
by the composition of the multiplication with projection. When $l
\to \infty$, this defines a deformation quantization of $\complex
\mathbb{P}^1$ (see \cite{bms:toeplitz}). The action of
$C^\infty(\complex \mathbb{P}^1)$ on $B_l$ defines a representation
of a fuzzy sphere. In the following we study the twisted Fock module
$\calf_m$ introduced in Proposition \ref{prop:tw-fock} in the case
of $\complex \mathbb{P}^1$. We relate the twisted Fock module
$\calf_m$ to the Berezin-Toeplitz quantization of $\complex
\mathbb{P}^1$.

We start by looking at the Fock module $\calf$. The module $\calf$
is the space of sections of an infinite dimensional vector bundle
over the manifold $X$ with fiber isomorphic to $p_\hbar \ast_\hbar
\calo_C$. This bundle is called the Fock bundle of $X$. In the next
two paragraphs, we construct a dense subbundle of the Fock bundle.

At each point $x\in X$, we look at the subspace $\calt^k _x$ of
the fiber $\calf_x$ of the Fock bundle, which is the span of
$p_\hbar \ast_\hbar y_{i_1}\ast_\hbar \cdots \ast_\hbar y_{i_k}$,
where $1\leq i_1, \cdots, i_k\leq n$. $\calt^k_x$ is a finite
dimensional subspace of $\calf_x$. As $p_\hbar \ast_\hbar
\bar{y}_i=0$, the direct sum of $\calt^k_x$ is equal to the
span of $p_\hbar\ast_\hbar f$ over
all polynomials $f(y, \bar{y})$. Since $p_\hbar \ast_\hbar
\calo_C$ is a subspace of $\cals$, the linear span of $p_\hbar
\ast_\hbar f(y, \bar{y})$ over all polynomials $f(y, \bar{y})$ is
dense in $p_\hbar \ast_\hbar \calo_C$. Therefore, $\oplus_k
\calt^k _x$ is dense in $\calf_x$.

Fixing $k$, we look at the collection of $\calt^k _x$ over $X$. This
is a finite dimensional vector bundle over $X$. The space of
sections of this bundle is denoted by $\calt^k$. According to the
above fiberwise discussion, on $\calt^k _x$, we know that $\Omega(X,
\oplus_k \calt^k)$ is dense in $\calf$. We notice that
$\nabla_\calf$ restricts to a connection on $\calt^k $ as a right
$C^\infty(X)$ module with the curvature equal to $-R/\hbar$. The
Chern character of $\calt^k$ is equal to
$Ch(\calt^k)=tr(\exp(-kR/\hbar))$. Using the fact that $p_\hbar
\ast_\hbar y_i ^{\ast k}\ast_\hbar (\bar{y_i}\ast_\hbar y_i)=k\hbar
(p_\hbar \ast_\hbar y_i^{\ast k})$, we compute the Chern character
of $\calt^k$ to be equal to $Ch(-TX)^k$, where $Ch(-TX)\in
H^{2\bullet}(X)$ is the Chern character of $-T X$.

In the following we look at a special case, $\complex \mathbb{P}^1$.
Since $\complex \mathbb{P}^1$ is one dimensional, $\calt^k$ is the
space of sections of a rank one bundle over $\complex \mathbb{P}^1$.
According to the above calculation of Chern character, we know that
$\calt^k$ corresponds to the bundle $K^{\otimes k}$, where $K$ is
the cotangent bundle of $\complex \mathbb{P}^1$. We conclude that in
the case of $\complex \mathbb{P}^1$, the Fock module $\calf$
contains a dense subspace $\calt$ which consists of sections of
\[
\bigoplus_{k\geq0} \Omega^*(\complex \mathbb{P}^1)\boxtimes
K^{\otimes k}.
\]

As a vector bundle over $\complex \mathbb{P}^1$, the module
$\calf_m$ can be viewed as the tensor of the Fock bundle $\calf$
with the line bundle $H_m$. According to the above analysis, we
know that $\calf_m$ contains a dense subspace $F_m$ which consists
of sections of
\[
\big(\bigoplus_{k\geq 0}\Omega^*(\complex \mathbb{P}^1)\boxtimes
K^{\otimes k}\big)\otimes H_m.
\]

In the case of $\complex \mathbb{P}^1$, the line bundle $H_m$ is the
m-th power of the dual bundle of the hyperplane bundle $H$ over
$\complex \mathbb{P}^1$ because its curvature is equal to
$-\textrm{im }\omega$. The canonical line bundle $K$ is dual to the
square of the hyperplane bundle $H$ because its curvature is equal
to $-2i\omega$. Therefore, the above dense subspace $F_m$ of the
twisted Fock module $\calf_m$ can be identified with
\[
\bigoplus_{k\geq 0}\Omega^*(\complex \mathbb{P}^1)\boxtimes
H^{\otimes (-m-2k)}.
\]

We conclude that the dual bundle of the twisted Fock module
$\calf_m$ contains the direct sum of bundles $H^{\otimes m}\oplus
H^{\otimes m+2}\oplus \cdots$.
\begin{prop}
The fuzzy sphere acts on the dual of the twisted Fock module
$\calf_m$ with the representation $\oplus_{k\geq0}B_{m+2k}$
\end{prop}

In the above, we introduced and studied the twisted Fock module over
the Q-algebra $(\Omega(X, \calw), \nabla, R)$, which is a special
example of quantization $dg-$modules. The twisted Fock module is an
infinitely generated projective module over $\Omega(X, \calw)$. One
must ask whether there are any finitely generated projective
quantization dg-modules.

The answer is affirmative in the case of a 2-dimensional torus
$\torus^2$ with a constant Poisson structure. According to
Proposition \ref{prop:quant}, the Q-algebra $\Omega(\torus^2,
\calw)$ is quasi-isomorphic to the algebra of the corresponding
quantum torus by the map $\calq:C^\infty(\torus^2)\to
\Omega(\torus^2, \calw)$. Given a projection $P$ of the quantum
torus, we define a right projective module $M_P$ of
$\Omega(\torus^2, \calw)$ by $\calq(P)\ast_\hbar \Omega(\torus^2,
\calw)$. It is straightforward to check that $M_P$ defines a
finitely generated projective quantization dg-module by the equation
$Q(\calq(P))=0$. Unfortunately, we do not know whether a general
symplectic manifold has any finitely generated projective
dg-modules. We end this note with the following conjecture:
\begin{conj}
For a symplectic manifold $M$, the Q-algebra $(\Omega(M, \calw),
\nabla, R)$ has finitely generated projective quantization
dg-modules. Furthermore, the K-group of finitely generated
projective quantization dg-modules over $(\Omega(M, \calw), \nabla,
R)$ is isomorphic to the K-group of finitely generated projective
modules over $C^\infty(M)$.
\end{conj}

\noindent{Department of Mathematics, University of California at Davis, Davis, CA, 95616, U.S.A.}\\
\noindent{xtang@math.ucdavis.edu}

\begin{thebibliography}{99}
\bibitem{aksz:bv}
Alexandrov, M., Kontsevich, M., Schwarz, A., and Zaboronsky, O.,
The geometry of the master equation and topological quantum field
theory, {\em Int. J. Mod. Phys.}, A12(1997), 1405-1403.

\bibitem{bms:toeplitz}
Bordemann, M., Meinrenken, E., and Schlichenmaier, M., Toeplitz
quantization of K\"ahler manifolds and ${\rm gl}(N)$, $N\to\infty$
limits, {\em Comm. Math. Phys.} 165 (1994), no. 2, 281--296.

\bibitem{cf:poi-sigma}
Cattaneo, A., and Felder, G., Poisson sigma models and deformation
quantization, {\em Mod. Phys. Lett.}, A16(2001), 179-190.

\bibitem{c:book}
Connes, A. {\em Noncommutative geometry. Academic Press}, Inc.,
San Diego, CA, 1994.

\bibitem{dkmm:smooth-algebra}
Dubois-Violette, M., Kriegl, A., Maeda, Y., and Michor, P., Smooth
$*$-algebras. {\em Noncommutative geometry and string theory
(Yokohama, 2001). Progr. Theoret. Phys. Suppl. } No. 144 (2001),
54--78.

\bibitem{fe:book}
Fedosov, B., {\em Deformation Quantization and Index Theory},
Mathematical Topics, 9, Akademie Verlag, Berlin, 1996.

\bibitem{gv:distribution}
Gracia-Bondia, J., V\'arilly, J., Algebras of distributions
suitable for phase-space quantum mechanics. I., {\em J. Math.
Phys. 29} (1988), no. 4, 869--879.

\bibitem{ka:sep-def}
Karabegov, A., On Fedosov's approach to deformation quantization
with separation of variables, Conf\'erence Mosh\'e Flato 1999, Vol.
II (Dijon), 167--176, {\em Math. Phys. Stud.}, 22, Kluwer Acad.
Publ., Dordrecht, 2000.

\bibitem{konts:deformation}
Kontsevich, M., Deformation quantization of Poisson manifolds,
{\em Lett. Math. Phys.} 66 (2003), no. 3, 157--216.

\bibitem{rif:quantization}
Rieffel, M., {\em Deformation quantization for actions of $\reals^
d$}, Mem. Amer. Math. Soc. 106 (1993), no. 506.



\bibitem{sc:bv-quant}
Schwarz, A., Geometry of Batalin-Vilkovisky quantization, {\em
Comm. Math. Phys.}, 155(1993), 249-260.

\bibitem{sc:semiclassical}
Schwarz, A., Semiclassical approximation in Batalin-Vilkovisky
Formalism, {\em Comm. Math. Phys.}, 158(1993), 375-396.

\bibitem{sc:ncsusy}
Schwarz, A., Noncommutative supergeometry and duality, {\em Lett.
Math. Phys.} 50 (1999), no. 4, 309--321.

\bibitem{sc:deformation}
Schwarz, A., Noncommutative supergeometry, duality and
deformations, {\em Nuclear Phys. B} 650 (2003), no. 3, 475--496.
\end{thebibliography}
\end{document}